\documentclass[11pt,a4paper]{article}
\pdfoutput=1
\usepackage{jcappub}
\usepackage[font=footnotesize,up]{caption}
\usepackage{subcaption}
\usepackage{wrapfig}
\usepackage{enumerate}
\usepackage{multirow}
\usepackage{booktabs}
\usepackage{bm}
\usepackage{array}
\usepackage{tabulary}
\usepackage{xspace}
\usepackage[normalem]{ulem}
\usepackage{lineno}
\usepackage{breakurl}
\usepackage{footnote}
\usepackage{tablefootnote}
\usepackage{threeparttable}
\usepackage{accents}

\newlength{\dhatheight}
\newcommand{\doublehat}[1]{%
    \settoheight{\dhatheight}{\ensuremath{\hat{#1}}}%
    \addtolength{\dhatheight}{-0.2ex}%
    \hat{\vphantom{\rule{1pt}{\dhatheight}}%
    \smash{\hat{#1}}}}

\newcommand{\svUL}{\langle \sigma \mathit{v} \rangle^\mathrm{UL}}
\newcommand{\sv}{\langle \sigma \mathit{v} \rangle}
\newcommand{\svsvt}{\sv_\mathrm{svt}}
\newcommand{\svdsu}{\sv_{2.71}}
\newcommand{\hatsv}{\widehat{\langle \sigma \mathit{v} \rangle}}
\newcommand{\doublehatnu}{\doublehat{\bm{\nu}}}
\newcommand{\hatnu}{\hat{\bm{\nu}}}
\newcommand{\hatmu}{\hat{\bm{\mu}}}

\newcommand{\data}{\bm{\mathcal{D}}}
\newcommand{\lkl}{\mathcal{L}}
\newcommand{\overbar}[1]{\mkern 3.5mu\overline{\mkern-3.5mu#1\mkern-3.5mu}\mkern 3.5mu}
\newcommand{\non}{{N_\mathrm{ON}}}
\newcommand{\noff}{{N_\mathrm{OFF}}}
\newcommand{\emin}{{E'_\mathrm{min}}}
\newcommand{\emax}{{E'_\mathrm{max}}}
\newcommand{\mdm}{m_\mathrm{DM}}
\newcommand{\dom}{{\Delta\Omega}}

\newcommand{\bb}{b\bar{b}}
\newcommand{\WW}{W^+W^-}
\newcommand{\tautau}{\tau^+\tau^-}
\newcommand{\mumu}{\mu^+\mu^-}
\newcommand{\Jobs}{J_\mathrm{obs}}
\newcommand{\Jobsi}{J_\mathrm{obs,i}}
\newcommand{\lp}{\lambda_P}
\newcommand{\Junits}{GeV$^2$ cm$^{-5}$}
\newcommand{\svunits}{cm$^3$ s$^{-1}$}
\newcommand{\degree}{\ensuremath{{}^{\circ}}\xspace}
\newcommand{\Fermi}{\textit{Fermi}\xspace}
\newcommand{\Tobs}{T_\mathrm{obs}}

\title{Limits to dark matter annihilation cross-section from a combined
  analysis of MAGIC and \Fermi-LAT observations of dwarf satellite
  galaxies}

\author [1] {M.~L.~Ahnen}
\author [2] {S.~Ansoldi}
\author [3] {L.~A.~Antonelli}
\author [4] {P.~Antoranz}
\author [5] {A.~Babic}
\author [6] {B.~Banerjee}
\author [7] {P.~Bangale}
\author [7,25] {U.~Barres de Almeida}
\author [8] {J.~A.~Barrio}
\author [9,26] {J.~Becerra Gonz\'alez}
% ...
\author{(the MAGIC Collaboration)}
\author[12,{\bf *}] {J.~Aleksi\'c}
\author[34,{\bf *}]{M.~Wood}
\author[35,36] {B.~Anderson}
\author[34] {E.~D.~Bloom}
\author[37] {J.~Cohen-Tanugi}
\author[38] {A.~Drlica-Wagner}
\author[39] {M.~N.~Mazziotta}
\author[35,36] {M.~S\'anchez-Conde}
\author[40] {L.~Strigari}

\affiliation [1] {ETH Zurich, CH-8093 Zurich, Switzerland}
\affiliation [2] {Universit\`a di Udine, and INFN Trieste, I-33100 Udine, Italy}
\affiliation [3] {INAF National Institute for Astrophysics, I-00136 Rome, Italy}
\affiliation [4] {Universit\`a  di Siena, and INFN Pisa, I-53100 Siena, Italy}
\affiliation [5] {Croatian MAGIC Consortium, Rudjer Boskovic Institute, University of Rijeka, University of Split and University of Zagreb, Croatia}
\affiliation [6] {Saha Institute of Nuclear Physics, 1/AF Bidhannagar, Salt Lake, Sector-1, Kolkata 700064, India}
\affiliation [7] {Max-Planck-Institut f\"ur Physik, D-80805 M\"unchen, Germany}
\affiliation [8] {Universidad Complutense, E-28040 Madrid, Spain}
\affiliation [9] {Inst. de Astrof\'isica de Canarias, E-38200 La Laguna, Tenerife, Spain; Universidad de La Laguna, Dpto. Astrof\'isica, E-38206 La Laguna, Tenerife, Spain}
% ...
\affiliation [12] {Institut de Fisica d'Altes Energies (IFAE), The Barcelona Institute of Science and Technology, Campus UAB, 08193 Bellaterra (Barcelona), Spain}
% ...
\affiliation [25] {now at Centro Brasileiro de Pesquisas F\'isicas (CBPF/MCTI), R. Dr. Xavier Sigaud, 150 - Urca, Rio de Janeiro - RJ, 22290-180, Brazil}
\affiliation [26] {now at NASA Goddard Space Flight Center, Greenbelt, MD 20771, USA and Department of Physics and Department of Astronomy, University of Maryland, College Park, MD 20742, USA}
% ...
\affiliation [34] {W. W. Hansen Experimental Physics Laboratory, Kavli Institute for Particle Astrophysics and Cosmology, Department of Physics and SLAC National Accelerator Laboratory, Stanford University, Stanford, CA 94305, USA}
\affiliation [35] {Department of Physics, Stockholm University, AlbaNova, SE-106 91 Stockholm, Sweden}
\affiliation [36] {The Oskar Klein Centre for Cosmoparticle Physics, AlbaNova, SE-106 91 Stockholm, Sweden}
\affiliation [37] {Laboratoire Univers et Particules de Montpellier, Universit\'e Montpellier, CNRS/IN2P3, Montpellier, France}
\affiliation [38] {Center for Particle Astrophysics, Fermi National Accelerator Laboratory, Batavia, IL 60510, USA}
\affiliation [39] {Istituto Nazionale di Fisica Nucleare, Sezione di Bari, I-70126 Bari, Italy}
\affiliation [40] {Texas A\&M University, Department of Physics and Astronomy, College Station, TX 77843-4242, USA}

\emailAdd{jrico@ifae.es}
\emailAdd{mdwood@slac.stanford.edu}
\emailAdd{jelena@ifae.es}

\abstract{

  We present the first joint analysis of gamma-ray data from the MAGIC
  Cherenkov telescopes and the {\it Fermi} Large Area Telescope (LAT) to
  search for gamma-ray signals from dark matter annihilation in dwarf
  satellite galaxies.
  We combine 158 hours of Segue~1 observations with MAGIC with 6-year
  observations of 15 dwarf satellite galaxies by the {\it
    Fermi}-LAT. We obtain limits on the annihilation cross-section for
  dark matter particle masses between 10 GeV and 100 TeV -- the widest
  mass range ever explored by a single gamma-ray analysis.  These
  limits improve on previously published \Fermi-LAT and MAGIC results by up
  to a factor of two at certain masses.  Our new inclusive analysis
  approach is completely generic
  and can be used to perform a global, sensitivity-optimized dark
  matter search by combining data from present and future gamma-ray
  and neutrino detectors.}


\begin{document}
\maketitle
\flushbottom

\section{Introduction}
\label{sec:intro}

The existence of a non-baryonic, neutral and cold dark matter (DM)
component in the Universe is supported by an overwhelming body of
observational evidence, mainly involving its gravitational effects on
the dynamics of cosmic structures like galaxy clusters (first observed
by Zwicky in 1933 \cite{ref:zwicky33}) and spiral galaxies
\cite{ref:babcock39,ref:salucci}, and on the power spectrum of
temperature anisotropies of the cosmic microwave background
\cite{ref:wmap,ref:planck2015}. Several theories beyond the Standard
Model postulate the existence of a new neutral, stable and weakly
interacting massive particle (generically known as
WIMP~\cite{ref:hut77}), with mass in the TeV scale, and that could
account for the measured DM relic density (e.g.~\cite{ref:gelmini20}).

A promising way to identify the nature of DM and to measure its
properties is to search for the Standard Model (SM) particles produced
through WIMP annihilation or decay at DM over-densities, or {\it halos}, in the
local Universe. Among those SM products, gamma rays and neutrinos are
the only stable neutral particles. They travel from their production
sites to Earth unaffected by magnetic deflection, and as such are
ideal messengers for astronomical DM searches.

Current gamma-ray instruments like the \Fermi Large Area Telescope
(LAT)\footnote{\url{http://fermi.gsfc.nasa.gov }} in space, the
ground-based Cherenkov telescopes
MAGIC\footnote{\url{http://magic.mpp.mpg.de}},
H.E.S.S.\footnote{\url{http://www.mpi-hd.mpg.de/hfm/HESS}} and
VERITAS\footnote{\url{http://veritas.sao.arizona.edu}}, the
new-generation water Cherenkov detector
HAWC\footnote{\url{http://www.hawc-observatory.org}}, as well as
neutrino telescopes like
Antares\footnote{\url{http://antares.in2p3.fr}},
IceCube\footnote{\url{http://icecube.wisc.edu}}, and
SuperKamiokande\footnote{\url{http://www-sk.icrr.u-tokyo.ac.jp/sk/index-e.html}},
are sensitive to overlapping and/or complementary DM particle mass
ranges (from $\sim$1 GeV to $\sim$100 TeV). All these instruments have
dedicated programs to look for WIMP signals coming from, e.g., the
Galactic Center and halo
\cite{hessGC,magicGC,veritasGC,hessGH,ref:FermiHalo,hessLines,ref:FermiLines,ref:hawcDM,ref:icecubeGH,ref:antaresDM},
galaxy clusters
\cite{magicPerseus,hessFornax,veritasComa,ref:FermiClusters,ref:icecubeClusters},
dwarf spheroidal galaxies (dSphs)
\cite{hessSgr,hessCMa,hessCarinaSculptor,hessDwarfs,veritasDwarfs,veritasSegueErr,magicDraco,magicWillman1,ref:MAGICSegue2012,ref:MAGICSegue2014,ref:Fermi2014,ref:Fermi2015}
and other targets \cite{ref:FermiIso,ref:FermiClumps,ref:icecubeSun}.

A natural step forward within this collective effort is the combination of data from different experiments and/or
observational targets, which allows a global and sensitivity-optimized search
\cite{ref:FullLikelihood}. This can be achieved in a relatively
straightforward way since, for a given DM particle model, a joint
likelihood function can be written as the product of the particular
likelihood functions for each of the measurements/instruments. One
advantage of such an approach is that the details of each
experiment, such as event lists or instrument response functions
(IRFs), do not need to be combined or averaged. 

In this paper, we present a new global analysis framework for DM
searches, applicable to observations from gamma-ray and neutrino
instruments, and the results of applying it to the MAGIC and
\Fermi-LAT observations of dSphs.

DSphs are associated with the population of Galactic DM sub-halos,
predicted by the cold DM structure formation scenario and reproduced
in N-body cosmological simulations
(e.g. \cite{ref:aquarius,ref:vialactea}), that have accreted enough
baryonic mass to form stars (other sub-halos may remain completely
dark).  DSphs have very high mass-to-light ratios, being the most
DM-dominated systems known so far \cite{ref:Strigari}.  They have the
advantage of being free of other sources of gamma-ray emission and
have been identified predominantly at high Galactic latitudes, where
diffuse astrophysical foregrounds from the Milky Way are lowest.
Because they are relatively close, they are expected to appear as
point-like or marginally extended sources for gamma-ray and neutrino
telescopes, with relatively high DM annihilation fluxes.  The stellar
kinematics of these systems can be used to determine their DM
distribution and its uncertainty using a common
methodology~\cite{ref:Martinez,ref:Geringer,ref:Bonnivard}. These
measurements enable the combination of dSph observations that
constrains models of DM annihilation or decay within the dSph
population as a whole.

This article is organized as follows: Section \ref{sec:instruments}
describes the MAGIC and \Fermi-LAT instruments, and the data sets used
in our study. The global analysis framework is described in
Section~\ref{sec:analysis}. The results of applying the analysis to
MAGIC and \Fermi-LAT data are presented in
Section~\ref{sec:results}. Finally, in Section~\ref{sec:discussion} we
discuss those results and present our conclusions.
 
\section{Instruments and Observations}
\label{sec:instruments}

\subsection{The MAGIC telescopes}
\label{sec:magic}

The \emph{Florian Goebel} Major Atmospheric Gamma-ray Imaging Cherenkov
(MAGIC) telescopes are located at the Roque de los Muchachos
Observatory (28.8\degree~N, 17.9\degree~W; 2200~m above sea level), on
the Canary Island of La Palma, Spain. MAGIC is a system of two
telescopes that detect Cherenkov light produced by the atmospheric
particle showers initiated by cosmic particles entering the Earth's
atmosphere. Images of these showers are projected by the MAGIC
reflectors onto the photo-multiplier tube (PMT) cameras, and are used
to reconstruct the calorimetric and spatial properties, as well as the
nature of the primary particle. Thanks to its large reflectors (17
meter diameter), plus its high-quantum-efficiency and low-noise PMTs,
MAGIC achieves high sensitivity to Cherenkov light and low energy
threshold~\cite{ref:MAGICupgrade2}. The MAGIC telescopes are able to
detect cosmic gamma rays in the very high energy (VHE) domain, i.e. in
the range between $\sim$50 GeV and $\sim$50 TeV.

For this work we use MAGIC data corresponding to 158 hours of
observations of the satellite galaxy Segue~1
\cite{ref:MAGICSegue2014}, the deepest observations of any dSph by a
Cherenkov telescope to date.\footnote{Older MAGIC single-telescope DM
  searches (Draco~\cite{magicDraco}, 7.8 hours;
  Willman~1~\cite{magicWillman1}, 15.5 hours; and
  Segue~1~\cite{ref:MAGICSegue2012}, 29.4 hours) yield a comparatively
  poor sensitivity~\cite{ref:MAGICSegue2014} and are
  therefore not included in this work.} The data were taken between
2011 and 2013, with observations before, during, and after a major
hardware upgrade~\cite{ref:MAGICupgrade1}. This resulted in the
necessity of defining 4 different observation periods, each described
by a different set of IRFs. Observations were performed in the
so-called wobble mode, which allows simultaneous observations of the
target and the background control regions. For that purpose, two
different positions $\sim$0.4\degree away from the position of
Segue~1 were tracked alternatively for 20 min runs. For each of these
positions, the spectral shape of the residual background is slightly
different, and needs to be modeled independently. Therefore, for
MAGIC, we consider 8 independent data sets, each consisting of the
gamma-ray candidate events plus the corresponding IRFs and residual
background models. The MAGIC data are analyzed with a one-dimensional
unbinned likelihood fit to the energy distribution within a signal (or
ON) region of radius 0.122\degree around the center of Segue~1
(optimized for a source with the angular size of Segue~1).  The
integral of the DM emission template within the ON region defines the
model for the expected signal distribution as a function of
energy. See Section \ref{sec:lkl} and Ref.\ \cite{ref:MAGICSegue2014}
for more details about the MAGIC data analysis.

\subsection{The \Fermi-LAT}

The \Fermi-LAT is a pair-conversion telescope that is sensitive to
gamma rays in the energy range from 20~MeV to more than 300~GeV
\cite{ref:Atwood2009}.  With its large field-of-view (2.4 sr), the LAT
is able to efficiently survey the entire sky.  Since its launch in
June 2008, the LAT has primarily operated in a survey observation
mode that continuously scans the entire sky every 3 hours.  The
survey-mode exposure coverage is fairly uniform with variations of at
most 30\% with respect to the average exposure.  The LAT point-source
sensitivity, which is dependent on the intensity of diffuse backgrounds,
shows larger variations but is relatively constant at high Galactic
latitudes ($|b| > 10$\degree).  More details on the on-orbit performance
of the LAT are provided in Ref.~\cite{Ackermann:2012kna}.

The \Fermi-LAT data set used in this work corresponds to 6 years of
observations of 15 dSphs. We analyzed the data with the latest (Pass
8) event-level analysis~\cite{ref:Fermi2015}, using the \Fermi Science
Tools version 10-01-01 and the P8R2\_SOURCE\_V6 IRFs. DM signal
morphological and spectral templates are used in a three-dimensional
likelihood fit to the distribution of events in reconstructed energy
and direction within the $10\degree\times10\degree$ dSph
region-of-interest (ROI).  The model for each ROI contains the dSph DM
intensity template, templates for Galactic and isotropic diffuse
backgrounds, and point sources taken from the third LAT source catalog
(3FGL) \cite{ref:Fermi3FGL} within 15\degree of the ROI center.  A
broadband fit in each ROI is performed fitting the normalizations of
the Galactic and isotropic diffuse components and 3FGL sources that
fall within the ROI boundary.  After performing the broadband fit, a
set of likelihoods is extracted for each energy bin by scanning the
flux normalization of a putative DM source modeled as a power law with
spectral index 2 at the location of the dSph.  Tables with likelihood
values versus energy flux for each energy bin are produced for all
considered targets.  The likelihood tables used for the present work
are taken from Ref.~\cite{ref:Fermi2015} and can be found in the
corresponding online
material.\footnote{\url{http://www-glast.stanford.edu/pub_data/1048/}.}
These tables allow the computation of joint-likelihood values for any
gamma-ray spectrum, and are used as input to the present analysis (see
Section \ref{sec:lkl} for more details).

\section{Analysis}
\label{sec:analysis}

\subsection{Dark Matter annihilation flux}

The gamma-ray (or neutrino) flux produced by DM annihilation in a
given region of the sky ($\dom$) and observed at Earth is given by:
\begin{equation}
\frac{d\Phi}{dE}(\dom) = \frac{1}{4\pi}\, \frac{\sv\, J(\dom)}{2\mdm^2}\,
\frac{dN}{dE}\quad ,
\label{eq:gammaflux}
\end{equation}
where $\mdm$ is the mass of the DM particle, $\sv$ the
thermally-averaged annihilation cross-section, $dN/dE$ the average
gamma-ray spectrum per annihilation reaction (for neutrinos this term
includes the oscillation probability between target and Earth), and
\begin{equation}
  J(\dom) = \int_{\dom}  d\Omega' \ \int_\mathrm{l.o.s.} dl\,
  \rho^2(l,\Omega')
\label{eq:Jfactor}
\end{equation}
is the so-called \emph{astrophysical factor} (or simply the J-factor),
with $\rho$ being the DM density, and the integrals running over
$\dom$ and the line-of-sight (l.o.s.)  through the DM distribution.

Using the PYTHIA simulation package version 8.205 \cite{ref:pythia},
we have computed the average gamma-ray spectrum per annihilation
process ($dN/dE$) for a set of DM particles of masses between 10 GeV
and 100 TeV (i.e.\ in the WIMP range), annihilating into SM pairs
$\bb$, $\WW$, $\tautau$ and $\mumu$. For each channel and mass, we
average the gamma-ray spectrum resulting from $10^7$ decay events of a
generic resonance with mass $2\times\mdm$ into the specified
pairs. For each simulated event, we trace all the decay chains,
including the muon radiative decay
($\mu^- \to e^- \bar{\nu}_e \nu_\mu \gamma$, not active in PYTHIA by
default), down to stable particles.

We take J-factors around the analyzed dSphs from Ackermann et al.\
\cite{ref:Fermi2015}, who follow the approach of
Ref.\ \cite{ref:Martinez}. The DM distributions in the halos of the dSphs are
parameterized following a Navarro-Frenk-White (NFW)
profile~\cite{ref:nfw}:
\begin{equation} 
\rho(r) = \frac{\rho_0 r^3_s}{r(r_s+r)^2}\quad,
\label{eq:rho}
\end{equation}
where $r_s$ and $\rho_0$ are the NFW scale radius and characteristic
density, respectively, and are determined from a fit to the dSph
stellar surface density and velocity dispersion profiles.
The properties of the dSphs used in our analysis, including the
J-factors and their uncertainties, are summarized in
Table~\ref{tab:dwarfs}. We quote the measured J-factors ($\Jobs$) for
a reference integrated radius of 0.5\degree from the halo center in
all cases, which encompasses more than 90\% of the annihilation flux
for our dSph halo models (which have halo scale radii between
0.1\degree and 0.4\degree). We note that Table~\ref{tab:dwarfs}
together with Equations \ref{eq:Jfactor} and \ref{eq:rho} allow the
computation of the J-factors for any other considered $\dom$, and
therefore the DM emission templates. In Ref.~\cite{ref:Fermi2015}
it has been shown how the {\it Fermi}-LAT limits can vary by up to
$\sim35\%$ (for a 100 GeV DM mass, and decreasing with the mass value)
by assuming different parameterizations for the DM density profile in
dSphs.

\begin{table}[t]
\begin{center}
\begin{tabular}{lrrrlc}
\hline
Name & l & b& D & $r_{s}/\mathrm{D}$ & $\log_{10}(\Jobs)$ \\
         &  [deg]  & [deg] & [kpc] & [deg] & $[\log_{10}(\mathrm{GeV}^2
         \mathrm{cm}^{-5})]$ \\
\hline
Bootes I          & 358.08 &  69.62 &  66 & 0.23  & 18.8 $\pm$ 0.22 \\
Canes Venatici II & 113.58 &  82.70 & 160 & 0.071 & 17.9 $\pm$ 0.25  \\
Carina            & 260.11 & -22.22 & 105 & 0.093 & 18.1 $\pm$ 0.23  \\
Coma Berenices    & 241.89 &  83.61 &  44 & 0.23  & 19.0 $\pm$ 0.25 \\
Draco             &  86.37 &  34.72 &  76 & 0.26  & 18.8 $\pm$ 0.16  \\
Fornax            & 237.10 & -65.65 & 147 & 0.17  & 18.2 $\pm$ 0.21 \\
Hercules          &  28.73 &  36.87 & 132 & 0.081 & 18.1 $\pm$ 0.25 \\
Leo II            & 220.17 &  67.23 & 233 & 0.071 & 17.6 $\pm$ 0.18 \\
Leo IV            & 265.44 &  56.51 & 154 & 0.072 & 17.9 $\pm$ 0.28 \\
Sculptor          & 287.53 & -83.16 &  86 & 0.25  & 18.6 $\pm$ 0.18 \\
Segue 1           & 220.48 &  50.43 &  23 & 0.36  & 19.5 $\pm$ 0.29 \\
Sextans           & 243.50 &  42.27 &  86 & 0.13  & 18.4 $\pm$ 0.27  \\
Ursa Major II     & 152.46 &  37.44 &  32 & 0.32  & 19.3 $\pm$ 0.28 \\
Ursa Minor        & 104.97 &  44.80 &  76 & 0.35  & 18.8 $\pm$ 0.19  \\
Willman 1         & 158.58 &  56.78 &  38 & 0.25  & 19.1 $\pm$ 0.31 \\
\hline
\end{tabular}
\caption{\label{tab:dwarfs} DSphs used in the present analysis and
  their main properties: Name, Galactic longitude and latitude,
  distance to Earth, angular size of the DM halo scale radius, and
  J-factor (with statistical uncertainty) assuming an NFW density
  profile and integrated to a radius of 0.5\degree from the dSph
  center.}\end{center}
\end{table}

Using Equation \ref{eq:gammaflux} together with the values of $dN/dE$
and $J$ obtained as detailed in the previous paragraphs, we compute
morphological and spectral intensity templates for the DM emission in
each dSph. Folding these templates with the response of the MAGIC and
LAT instruments, we compute the expected count distribution as a
function of the measured energy and position within the observed field
of each dSph.

\subsection{Likelihood analysis}
\label{sec:lkl}

For each considered annihilation channel and DM particle
mass, we compute the profile likelihood ratio as a function of $\sv$
(see, e.g.\ \cite{ref:pdg}):
\begin{equation}
\lp(\sv\, |\, \data) = \frac{\lkl(\sv; \doublehatnu\,
  |\, \data)}{\lkl(\hatsv; \hatnu\,  |\, \data)}\quad ,
\label{eq:profile} 
\end{equation}
with $\data$ representing the data set and $\bm{\nu}$ the nuisance
parameters.\footnote{In statistics, nuisance parameters are those that
  are not of intrinsic interest but that must be included for an
  accurate description of the data.} $\hatsv$ and $\hatnu$ are the
values maximizing the joint likelihood function ($\lkl$), and
$\doublehatnu$ the value that maximizes $\lkl$ for a given value of
$\sv$. The likelihood function can be written as:
\begin{equation}
\lkl (\sv;\bm{\nu}\, |\, \data) =
\prod_{i=1}^{N_\mathrm{target}}  \lkl_i (\sv;  J_i,\bm{\mu}_i\, |\, 
  \data_i) \cdot \mathcal{J}(J_i\, |\, J_{\mathrm{obs},i},\sigma_i)\quad ,
\label{eq:lkl-target}
\end{equation}
where the index $i$ runs over the different targets (dSphs in our
case); $J_i$ is the J-factor for the corresponding target (see
Equations \ref{eq:Jfactor} and \ref{eq:rho}); $\bm{\mu}_i$ denotes any
nuisance parameters additional to $J_i$; and $\data_i$ is the
target-related input data.  $\mathcal{J}$ is the likelihood for $J_i$,
given measured $\log_{10}(\Jobsi)$ and its uncertainty $\sigma_i$
\cite{ref:Fermi2015}:
\begin{equation}
  \mathcal{J}(J_i\, |\, \Jobsi,\sigma_i) = \frac{1}{\ln(10) \Jobsi \sqrt{2\pi}\sigma_i}
  \times
  e^{-\big(\log_{10}(J_i)-\log_{10}(\Jobsi)\big)^2/2\sigma_i^2}\quad.
\label{eq:jfactorPDF}
\end{equation}
The likelihood function for a particular target ($\lkl_i$) can in turn
be written as the product of the likelihoods for different instruments
(represented by the index $j$), i.e.:
\begin{equation}
  \lkl_i (\sv;  J_i,\bm{\mu}_i\, |\, \data_i) =
  \prod_{j=1}^{N_\mathrm{instrument}} \lkl_{ij}(\sv; J_i, 
  \bm{\mu}_{ij}\, |\, \data_{ij})\quad ,
\label{eq:lkl-instrument}
\end{equation}
where $\bm{\mu}_{ij}$ and $\data_{ij}$ represent the nuisance
parameters and input data set for the given target $i$ and instrument
$j$. Equations \ref{eq:profile}, \ref{eq:lkl-target} and
\ref{eq:lkl-instrument} are generic, i.e.\ they are valid for any set
of instruments and observed targets.\footnote{It is also worth noting
  that the values of $\lkl_{ij}$ ultimately depend on the flux of
  DM-induced gamma rays, hence on the product $\sv \cdot J_i $ (see
  Equation~\ref{eq:gammaflux}). Therefore, in order to compute
  $\lkl_i$ and $\lkl$ (and its profile with respect to $J_i$) it is
  enough to know (in addition to $\mathcal{J}$) the values of
  $\lkl_{ij}$ vs.\ $\sv$, for fixed values of $J_i$ (e.g.\ for
  $J_{\mathrm{obs},i}$), since:
\begin{equation} \lkl_{ij}(\sv; J_i, 
  \bm{\mu}_{ij}\, |\, \data_{ij}) = \lkl_{ij}\big(\frac{J_i}{J_{\mathrm{obs},i}} \sv,
  J_{\mathrm{obs},i}, \bm{\mu}_{ij}\, |\,
  \data_{ij}\big) \quad .
\label{eq:prop}
\end{equation}
This is a particularly useful property, since it allows a significant reduction
of the computing time requested for the profile of $\lkl$
over $J_i$, which can be explicitly written as:
\begin{equation}
\lkl(\sv; \hatnu\, |\, \data) =
\prod_{i=1}^{N_\mathrm{target}} \mathrm{max}_{J_i} \{\lkl_i(\sv;
  J_i, \hatmu\, |\, \data_i) \cdot \mathcal{J}(J_i\, |\,
  J_{\mathrm{obs},i}, \sigma_i) \}\quad ,
\end{equation}
where the values of $\lkl_i$ vs.\ $J_i$ can be computed using
Equations~\ref{eq:lkl-instrument} and \ref{eq:prop}. In addition, this
allows the combination of results from different instruments and
targets, starting from tabulated values of $\lkl_{ij}$ vs.\ $\sv$, for a
fixed value of $J_i$ and profiled with respect to
$\bm{\mu}_{ij}$. These values can be produced and shared by different
experiments without the need of releasing or sharing any of the
internal information used to produce them, such as event lists or IRFs.}

In our case, the likelihood for a given target $i$ observed by the
\Fermi-LAT ($j\equiv F$) is computed as:
\begin{equation}
\lkl_{iF}(\sv; J_i, \hatmu_{iF}\, |\, \data_{iF}) =
\prod_{k=1}^{N_\mathrm{E-bins}} \lkl_{iFk} \big(\overbar{E\Phi}_k (\sv,  J_i)\big) \quad ,
\end{equation}
with $k$ running over energy bins, and
\begin{equation}
\overbar{E\Phi}_k(\sv,  J_i) = \int_{E_{\mathrm{min},k}}^{E_{\mathrm{max},k}}
dE\, E\, \frac{d\Phi}{dE}(\sv, J_i) \quad .
\end{equation}
The values of $\lkl_{iFk}$ vs.\ $\overbar{E\Phi}$ corresponding to 6
years of observations of each of the considered dSph are tabulated and
released by the \Fermi-LAT
Collaboration~\cite{ref:Fermi2014,ref:Fermi2015}.

For the case of MAGIC ($j\equiv M$), the likelihood corresponding to a given
target $i$ can be written as:
\begin{equation}
\lkl_{iM} (\sv; J_i, \bm{\mu}_{iM}\, |\, \data_{iM}) = 
\prod_{k=1}^{N} \lkl_{iMk}(\sv; J_i, 
\bm{\mu}_{iMk}\, |\, \data_{iMk}) \quad ,
\label{eq:lklmagic}
\end{equation}
with the index $k$ running over $N=8$ different data sets (each described
by a different IRF, see Section~\ref{sec:magic}). The likelihood for a
given data set follows the method described in Refs.\
\cite{ref:FullLikelihood} and \cite{ref:MAGICSegue2014} and can be
written as (target, experiment and data set indices are omitted for the
sake of clarity):
\begin{eqnarray}
\lkl(\sv; J, \bm{\mu}\, |\, \data) & =  &\lkl\big(g(\sv, J,); b,
\tau\, |\, \{E'_l\}_{l=1,\dots, \non}, \{E'_m\}_{m=1,\dots, \noff}\big)
\\ 
 &  = & \frac{(g+b/\tau)^\non}{\non!} e^{-(g+b/\tau)}\,
 \frac{b^\noff}{\noff!} e^{-b}\, \prod_{l=1}^{\non} f( g; b, \tau\, |\,E'_l) \prod_{m=1}^{\noff} h(E'_m) \quad , \nonumber
\end{eqnarray}
where $g$ is the expected number of gamma rays detected with
reconstructed energy $E'$ in the telescope range $[\emin,\emax]$ and
an observation time $\Tobs$, i.e.:
\begin{equation}
g(\sv, J) = \Tobs \int_\emin^\emax dE' \int_0^\infty dE
\frac{d\Phi}{dE}(\sv, J)\, A(E)\, G(E;E') \quad ,
\end{equation}
$\tau$ and $b$ (nuisance parameters)
are the ratio of exposures between the OFF (background control) and ON
(signal) regions and the expected number of background events in the
OFF region, respectively. $h$ and $f$ are, respectively, the probability density
functions (PDFs) for measured OFF and ON events
with reconstructed energy $E'$. $h$ is obtained by fitting or
interpolating the measured differential event rate from one or several
additional background control regions observed simultaneously to the
ON and OFF regions (see Ref.\ \cite{ref:AleksicThesis} for
details). Finally, $f$ can be written as:
\begin{equation}
  f(g; b, \tau \, |\, E') = \frac{\frac{b}{\tau} h(E') + g\,
  p(E')}{\frac{b}{\tau} + g}\quad ,
\end{equation}
with
\begin{equation}
p(E') = \frac{\Tobs \int_0^\infty dE \frac{d\Phi}{dE}(\sv, J)\, A(E)\,
  G(E;E')}{g}\quad .
\label{eq:pdfp}
\end{equation}
$A(E)$ is the telescope effective area computed for a gamma-ray source
with the morphology expected for Segue~1 according to
Equations~\ref{eq:gammaflux} to \ref{eq:rho}, after analysis cuts
(including the selection of events in the ON region with radius of
0.122\degree around Segue~1 center). $G(E;E')$ is the PDF for the
energy estimator ($E'$) for a given true energy ($E$). Note that $p$
is also a PDF and therefore does not depend on $\Tobs$, $J$ or $\sv$.

\section{Results}
\label{sec:results}

We compute one-sided, 95\% confidence level upper limits to $\sv$ by
numerically solving the equation
$-2\ln \lp(\svdsu\, |\, \data) = 2.71$, with $\sv$ restricted to the
physical ($\geq$0) region. This prescription is the one used by the
\Fermi-LAT Collaboration in Refs.\ \cite{ref:Fermi2014,ref:Fermi2015},
and differs slightly from the one used by MAGIC in Ref.\
\cite{ref:MAGICSegue2014}. This has consequences on the comparison of
the results presented here with previous MAGIC results, which will be
discussed below.

\begin{figure}[tbp]
\centering
\includegraphics[width=.5\textwidth]{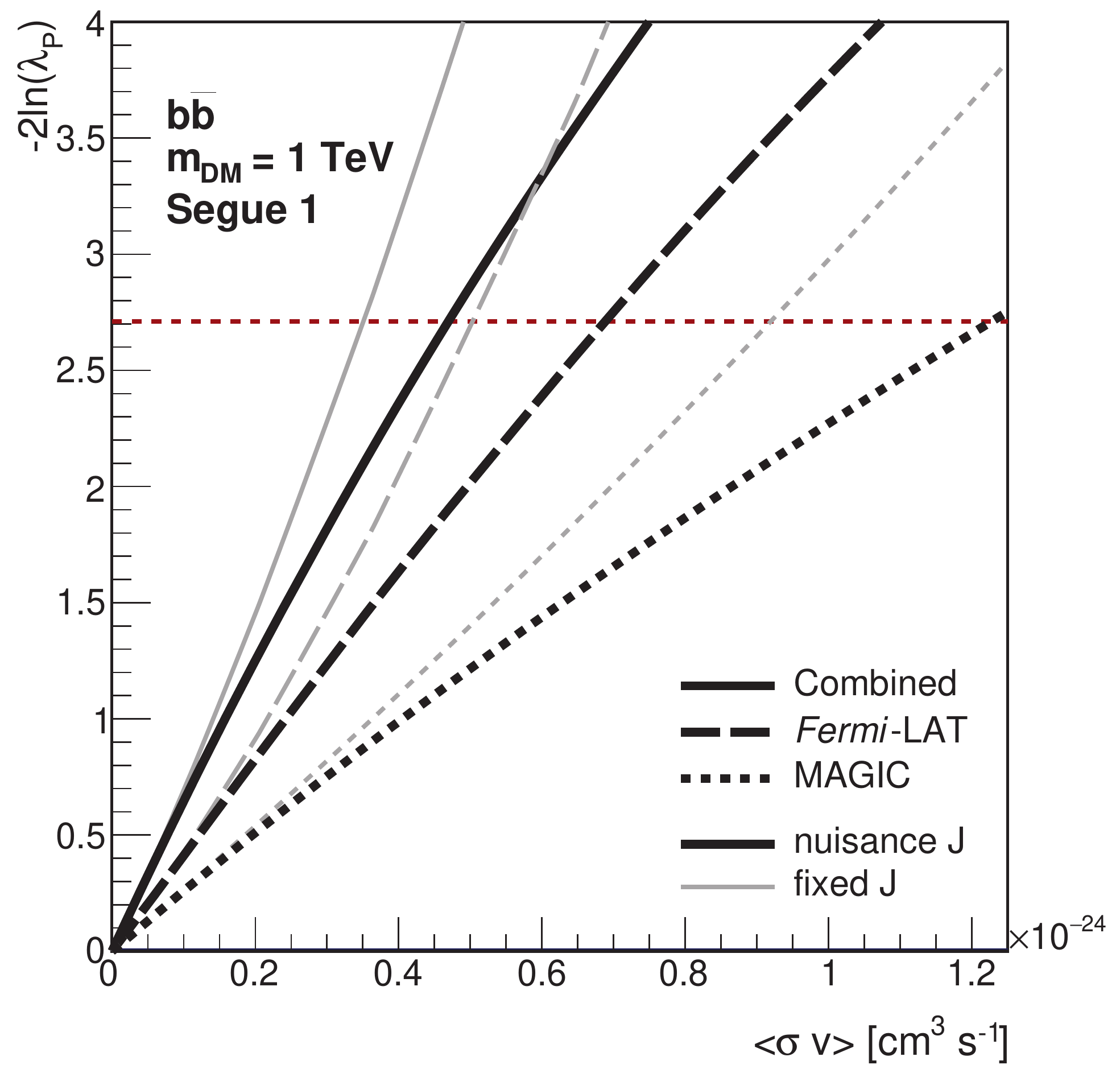}
\hfill
\caption{\label{fig:lklVSsv} $-2\ln(\lp)$ vs.\ thermally-averaged cross-section ($\sv$) for
  1~TeV DM particles annihilating into $\bb$ pairs in the
  Segue~1 dSph. Dotted, dashed and solid lines
  represent the values for MAGIC, \Fermi-LAT, and their
  combination, respectively. Thick-black and thin-gray lines
  represent the $-2\ln(\lp)$ functions when the J-factor is
  considered as a nuisance or fixed parameter, respectively. The
  horizontal dashed-red line shows the level determining, for each
  case, the one-sided 95\% confidence level upper limits.}
\end{figure}

Figure~\ref{fig:lklVSsv} illustrates for an individual target how the
combination of experiments improves the sensitivity to DM searches,
and how the J-factor statistical uncertainties worsen it. The figure
shows $-2\ln(\lp)$ vs.\ $\sv$ for $\mdm=1$\,TeV DM
particles annihilating into $\bb$ pairs in Segue~1. MAGIC and
\Fermi-LAT individual curves are shown, as well as their combination.
The effect of the statistical uncertainty on the J-factor is also
illustrated by showing curves where $J$ is treated as either a
nuisance or fixed parameter.  In each case, the 95\% confidence level
upper limits are obtained from the crossing point between the
corresponding curve and the $-2\ln(\lp) = 2.71$ line.  In the case of
treating $J$ as a nuisance parameter, MAGIC and \Fermi-LAT individual
limits are $\svUL=1.2\times10^{-24}$ \svunits\ and $6.8\times10^{-25}$
\svunits, respectively, whereas the combined analysis yields
$\svUL=4.6\times 10^{-25}$ \svunits.  Considering no uncertainties in
$J$ produces the combined limit $\svUL = 3.4\times 10^{-25}$ \svunits.

\begin{figure}[tbp]
\centering
\includegraphics[width=1\textwidth]{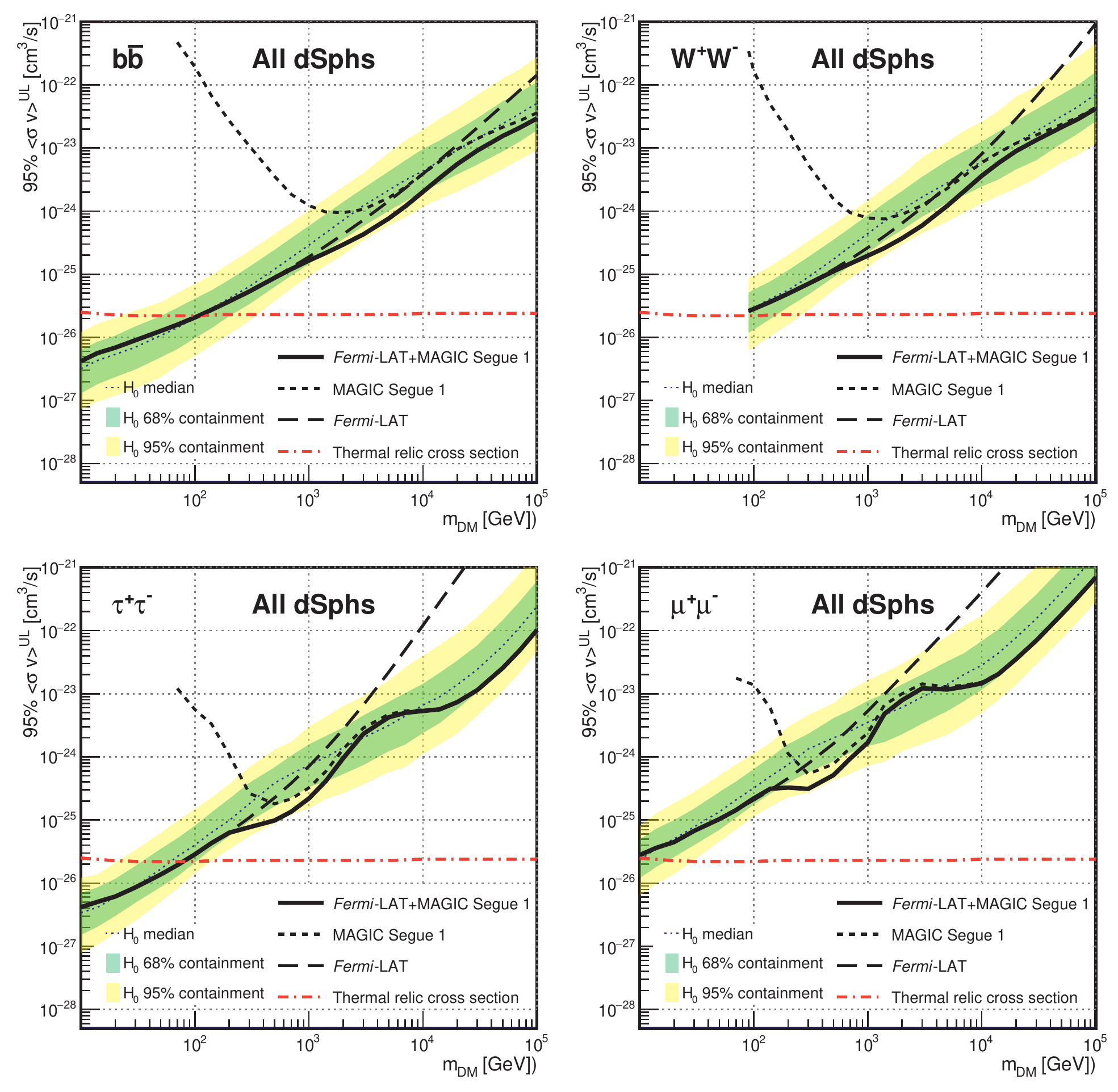}
\hfill
\caption{\label{fig:alldwarfs} 95\% CL upper limits on the
  thermally-averaged cross-section for DM particles annihilating into
  $\bb$ (upper-left), $W^+W^-$ (upper-right), $\tautau$ (bottom-left)
  and $\mu^+\mu^-$ (bottom-right) pairs. Thick solid lines show the
  limits obtained by combining {\it Fermi}-LAT observations of 15
  dSphs with MAGIC observations of Segue~1. Dashed lines show the
  observed individual MAGIC (short dashes) and {\it Fermi}-LAT (long dashes)
  limits. J-factor statistical uncertainties (Table \ref{tab:dwarfs})
  are considered as described in Section~\ref{sec:lkl}. The
  thin-dotted line, green and yellow bands show, respectively, the
  median and the symmetrical, two-sided 68\% and 95\% containment
  bands for the distribution of limits under the null hypothesis (see
  main text for more details). The red-dashed-dotted line shows the
  thermal relic cross-section from Ref.\ \cite{ref:Steigman2012}.}
\end{figure}

\begin{figure}[tbp]
\centering
\includegraphics[width=1\textwidth]{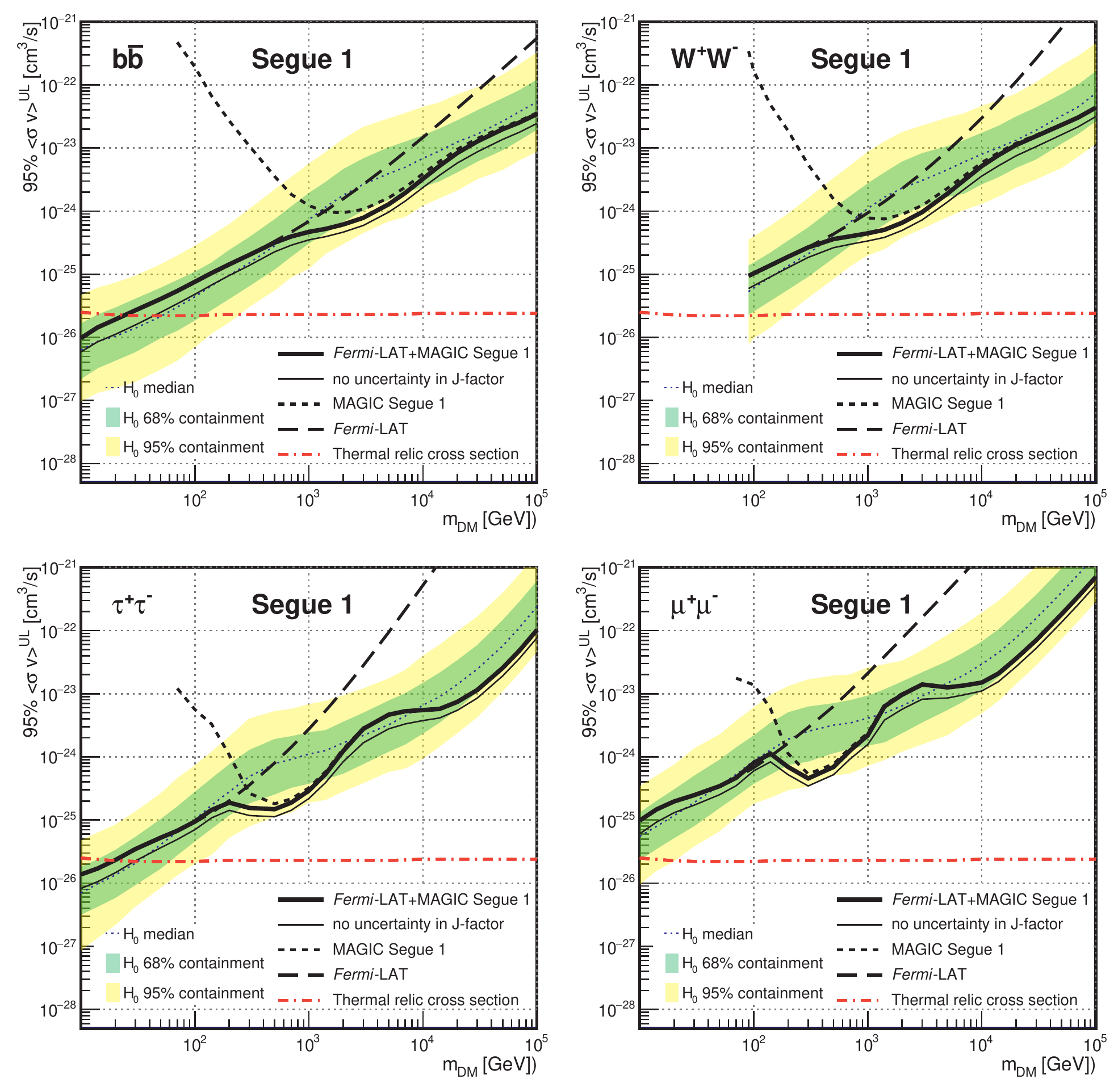}
\hfill
\caption{\label{fig:segue1} Similar as Figure~\ref{fig:alldwarfs} using
  Segue~1 observations only. The combined limits for the case when the
J-factor is considered as a fixed (no statistical uncertainties)
parameter are shown as a thin-solid line.}
\end{figure}

Figures~\ref{fig:alldwarfs} and \ref{fig:segue1} show our main
results: the 95\% confidence level limits to $\sv$ for DM particles
with masses between 10 GeV and 100 TeV annihilating into SM pairs
$\bb$, $\WW$, $\tautau$ and $\mumu$, as obtained from the combination
of \Fermi-LAT and MAGIC observations of 15 dSphs and Segue~1 alone,
respectively. We also show the observed limits obtained with MAGIC and
\Fermi-LAT data considered individually. The combined limits are
compared to the median, and the 68\% and 95\% containment bands,
expected under the null ($H_0 : \sv=0$) hypothesis. These are
estimated from the distributions of limits obtained by applying our
analysis to 300 independent $H_0$ realizations. Each realization
consists of data from \Fermi-LAT observations of one empty field per
considered dSph, combined with fast simulations of MAGIC Segue~1
observations. The empty fields constituting the LAT realizations were
selected by choosing random sky positions with $|b| > 30$\degree
centered at least $0.5$\degree away from a source in the 3FGL catalog.
MAGIC fast simulations consist of a set of event energies randomly
generated according to the background PDF, $h$ (see
Section~\ref{sec:lkl} and Ref.~\cite{ref:MAGICSegue2014} for details),
for both ON and OFF regions. In all cases, we assume the same
exposures on the different considered dSph as for the real data, and
J-factors randomly selected according to the PDF in
Equation~\ref{eq:jfactorPDF}.

We observe no evidence for DM annihilation in our data set (which
would appear as a positive deviation of the limit from the null
hypothesis). The maximum such deviation has a (local) significance of
about $0.3\,\sigma$, found for $\mdm \simeq 10$ GeV in the $\bb$ and
$\tautau$ annihilation channels, and for $\mdm \simeq 3$ TeV in the
$\mumu$ and $\tautau$ annihilation channels.  We also observe a
negative $2\,\sigma$ fluctuation at $\mdm \simeq 5$ TeV in the $\bb$
and $\WW$ annihilation channels and $\mdm \simeq 500$ GeV in the
$\mumu$ and $\tautau$ annihilation channels.  A deviation of this
magnitude would be expected in 5\% of the experiments under the null
hypothesis and is therefore compatible with random fluctuations.

As expected, limits in the low and high ends of the considered mass
range are dominated by \Fermi-LAT and MAGIC observations,
respectively, and the combined limits coincide with the individual
ones. The combination provides a significant improvement in the range
between $\sim$1 and $\sim$100 TeV (for $\bb$ and $\WW$) or $\sim$0.2
and $\sim$2 TeV (for $\tautau$ and $\mumu$), with a maximum
improvement of the combined limits with respect to the individual ones
by a factor $\sim$2. MAGIC individual limits shown in this work are
stronger by up to a factor $\sim$4 than those presented in Ref.\
\cite{ref:MAGICSegue2014}, which needs a dedicated explanation,
provided in Appendix~\ref{app:MAGICcomp}.

Systematic uncertainties in the determination of the J-factors would
weaken the limits on $\sv$ that can be inferred from the MAGIC and
\Fermi-LAT observations.  In this work we take the J-factors and
associated uncertainties from Ref.~\cite{ref:Martinez}, which are
largely consistent with the independent analysis of
Ref.~\cite{ref:Geringer}.  The analysis presented in
Ref.~\cite{ref:Bonnivard}, using a more flexible parameterization for
the stellar velocity distributions and more stringent criteria for
stellar membership, produces substantially larger J-factor
uncertainties for the ultra-faint dSphs used in the present study
(0.4--2.0 dex versus 0.2--0.3 dex).  They find the J-factor of
Segue~1 to be particularly uncertain due to the ambiguous membership
status of a few stars, and when excluding stars with membership
probability less than 95\% they find $\log_{10}(\Jobs$/(\Junits)$) =
$17.0$^{+2.1}_{-2.2}$.  This value is more than two orders of
magnitude smaller than the J-factor used in the present analysis
($\log_{10}(\Jobs$/(\Junits)$) = $19.5$\pm 0.3$) and would imply substantially
weaker limits from the MAGIC observations of Segue~1.  Although a full
examination of the J-factor uncertainties is beyond the scope of the
present work, we note that these uncertainties do not diminish the
power of the joint likelihood approach and can be fully accounted for
in our analysis procedure by updating the uncertainty parameter in the
J-factor likelihood (Equation \ref{eq:jfactorPDF}).

\section{Discussion and Conclusions}
\label{sec:discussion}

This work presents, for the first time, limits on the DM annihilation
cross-section from a comprehensive analysis of gamma-ray data with
energies between 500 MeV and 10 TeV. Using a common analysis approach
(both in the applied statistical methods and in the determination of
the J-factors) we have combined the MAGIC observations of Segue~1 with
{\it Fermi}-LAT observations of 15 dSphs. This allowed the computation
of meaningful global DM limits, and the direct comparison of the
individual results obtained with the different instruments. Our
results span the DM particle mass range from 10 GeV to 100 TeV -- the
widest range covered by a single gamma-ray analysis to date.

We find no signal of DM in our data set. Consequently, we set upper
limits to the annihilation cross-section. The obtained results are the
most constraining bounds in the considered mass range, from
observations of dSphs. For the low-mass range, our results (dominated
by \Fermi-LAT) are below the thermal relic cross-section $\sv \simeq
2.2\times10^{-26} $ \svunits. In the intermediate mass range (from few
hundred GeV to few tens TeV, depending on the considered annihilation
channel), where \Fermi-LAT and MAGIC achieve comparable sensitivities,
the improvement of the combined result with respect to the individual
ones reaches a factor $\sim$2. In addition, we present limits to DM
particle masses above 10 TeV (dominated by MAGIC) that have not been
shown before.

Our global analysis method is completely generic, and can be easily
extended to include data from more targets, instruments and/or
messenger particles provided they have similar sensitivity to the
considered DM particle mass range. Of particular interest is the case
of a global DM search from dSphs including data from all current
gamma-ray (\Fermi-LAT, MAGIC, H.E.S.S, VERITAS, HAWC) and neutrino
(Antares, IceCube, SuperKamiokande) instruments, and we hereby propose
a coordinated effort toward that end. Including results obtained from
other types of observational targets like the Galactic Center, galaxy
clusters or others is formally also possible, but a common approach to
the J-factor determination remains an open question. In the future,
this analysis could include new instruments like CTA~\cite{ref:ctaDM},
GAMMA-400~\cite{ref:gamma400DM}, DAMPE~\cite{ref:dampe} or
Km3Net~\cite{ref:km3netDM}.\footnote{The combination with results from
  direct searches or accelerator experiments following a similar
  approach is, in principle, also formally possible, but it would
  necessarily be model-dependent. This possibility should be however
  regarded as the culminating step of our proposal.} Our global
approach offers the best chances for indirect DM discovery, or for
setting the most stringent limits attainable by these kind of
observations.

\acknowledgments

The MAGIC Collaboration thanks
the Instituto de Astrof\'{\i}sica de Canarias
for the excellent working conditions
at the Observatorio del Roque de los Muchachos in La Palma.
The financial support of the German BMBF and MPG,
the Italian INFN and INAF,
the Swiss National Fund SNF,
the ERDF under the Spanish MINECO (FPA2012-39502), and
the Japanese JSPS and MEXT
is gratefully acknowledged.
This work was also supported
by the Centro de Excelencia Severo Ochoa SEV-2012-0234, CPAN CSD2007-00042, and MultiDark CSD2009-00064 projects of the Spanish Consolider-Ingenio 2010 programme,
by grant 268740 of the Academy of Finland,
by the Croatian Science Foundation (HrZZ) Project 09/176 and the University of Rijeka Project 13.12.1.3.02,
by the DFG Collaborative Research Centers SFB823/C4 and SFB876/C3,
and by the Polish MNiSzW grant 745/N-HESS-MAGIC/2010/0.

The \textit{Fermi} LAT Collaboration acknowledges generous ongoing
support from a number of agencies and institutes that have supported
both the development and the operation of the LAT as well as
scientific data analysis.  These include the National Aeronautics and
Space Administration and the Department of Energy in the United
States, the Commissariat \`a l'Energie Atomique and the Centre
National de la Recherche Scientifique / Institut National de Physique
Nucl\'eaire et de Physique des Particules in France, the Agenzia
Spaziale Italiana and the Istituto Nazionale di Fisica Nucleare in
Italy, the Ministry of Education, Culture, Sports, Science and
Technology (MEXT), High Energy Accelerator Research Organization (KEK)
and Japan Aerospace Exploration Agency (JAXA) in Japan, and the
K.~A.~Wallenberg Foundation, the Swedish Research Council and the
Swedish National Space Board in Sweden.
 
Additional support for science analysis during the operations phase is
gratefully acknowledged from the Istituto Nazionale di Astrofisica in
Italy and the Centre National d'\'Etudes Spatiales in France.

\appendix
\section{Comparison with previous MAGIC results}
\label{app:MAGICcomp}

The data, IRFs and likelihood functions (Equations \ref{eq:lklmagic}
to \ref{eq:pdfp}) used in this work are the same as for previous MAGIC
Segue~1 publication~\cite{ref:MAGICSegue2014}. Aside from enlarging
the explored DM mass range, the only differences between the two works
are: the assumed J-factor central value, the treatment of the J-factor
statistical uncertainties, and the prescription used for cases when
the maximum of the profile likelihood is found in the non-physical
($\sv<0$) region. We motivate each of these changes and comment on
their effect in the following paragraphs.

\begin{itemize}

\item In this work, J-factors are obtained following Ref.\
  \cite{ref:Martinez} and assuming an NFW DM density profile, whereas
  previous MAGIC results were obtained assuming an Einasto profile
  \cite{ref:MAGICSegue2012,ref:Essig2010}. This change has been
  introduced to homogenize MAGIC and \Fermi-LAT computation of
  J-factors, and produces a factor 2 lower (stronger) MAGIC limits with
  respect to the previously published ones.

\item Previous MAGIC results did not include statistical uncertainties
  in the J-factor. This was justified by the fact that $\svUL$ scales
  with $1/J$, and therefore the provided results allow to compute
  limits for any J-factor value. This argument is true only for
  single-target limits, but not for results obtained combining
  observations from different targets with different $J$ values and
  uncertainties. In this study, we include the statistical
  uncertainties on $J$ for all targets as described in Section
  \ref{sec:lkl}. In consequence, MAGIC limits increase (degrade) by a
  factor between $\sim$1.4 and $\sim$1.8 (depending on the considered
  annihilation channel and mass) with respect to the previously
  published ones.

  However, for the scalability argument given before, it is also
  interesting to provide single-target (Segue~1) results with and without
  uncertainties on the J-factor, which are shown in
  Figure~\ref{fig:segue1}.

\item The requirement of producing upper limits in the physical ($\sv
  \geq 0$) region normally leads to ad-hoc recipes implying
  over-coverage (see e.g.\ Ref.\ \cite{ref:pdg}). The problem arises
  when $\hatsv<0$, and it is aggravated when $\svdsu<0$. One possible
  prescription to deal with these cases is to restrict
  $-2\ln \lp(\sv\, |\, \data)$ to $\sv \geq 0$ \cite{ref:Rolke2005}. This is
  the solution adopted by \Fermi-LAT in Refs.\
  \cite{ref:Fermi2014,ref:Fermi2015}, and the one we have followed in
  this work. 

  Another, more conservative (i.e.\ producing larger over-coverage)
  prescription, was followed by MAGIC in Ref.\
  \cite{ref:MAGICSegue2014}, consisting in computing the 95\%
  confidence limits as $\svsvt=\svdsu-\hatsv$, whenever
  $\hatsv<0$. $\svsvt$ corresponds to what the authors in Ref.\
  \cite{ref:FeldmanAndCousins} define as the ``sensitivity'' of the
  measurement. Using this prescription, $\svsvt$ is the lowest
  possible value of the upper limit, irrespective of the presence of
  arbitrarily intense negative background fluctuations.

  The latter approach cannot be applied to \Fermi-LAT analysis since,
  due to low bin statistics, the likelihood function can be undefined
  for negative $\sv$ values (see Equation 4 in Ref.\
  \cite{ref:Fermi2015}). We therefore homogenize the treatment of
  limits close to the physical boundary for our whole data set by
  limiting $\sv$ to non-negative values. The difference between old
  and new prescriptions in MAGIC individual limits goes, depending on
  the annihilation channel and DM particle mass, from none (when
  $\hatsv \geq 0$) up to a factor $\sim$4 (for $\sim$2$\sigma$
  background fluctuations, see Figures~\ref{fig:alldwarfs} and
  \ref{fig:segue1}).
\end{itemize}

\end{document}